\documentclass[a4paper,11pt,aps,preprintnumbers,amsmath,amssymb,superscriptaddress,nofootinbib]{article}
\pdfoutput=1
\usepackage{jcappub}

\usepackage{multirow}
\usepackage{amssymb}
\usepackage{enumerate}
\usepackage{amssymb}
\usepackage{amsmath}
\usepackage{tikz}
\usepackage{feynmf}
\usepackage{makecell}
\usepackage{slashed}
\usepackage{natbib}
\usepackage{graphicx}
\hypersetup{colorlinks=true}

\usepackage{mathrsfs,amssymb}  
\usepackage{cancel}
\usepackage[normalem]{ulem}

\definecolor{darkgreen}{rgb}{0.01, 0.75, 0.24}
\definecolor{darkorange}{rgb}{1.0, 0.55, 0.0}
\usepackage{pifont}
\newcommand{\comment}[1]{}
\newcommand{\beq}{\begin{equation}}
\newcommand{\eeq}{\end{equation}}
\newcommand\bea{\begin{eqnarray}}
\newcommand\eea{\end{eqnarray}}

\newcommand{\eq}[1]{Eq.~(\ref{#1})}
\newcommand{\eqs}[2]{Eqs.~(\ref{#1}) and (\ref{#2})}

\renewcommand{\l}{\left}
\renewcommand{\r}{\right}

\title{Supersymmetric $U(1)_{B-L}$ flat direction and NANOGrav 15 year data}

\author[a]{Rinku Maji}
\author[b,c,d]{and Wan-il Park}
\note{\color{blue}All authors contributed equally.}
\affiliation[a]{
Laboratory for Symmetry and Structure of the Universe, Department of Physics, Jeonbuk National University, Jeonju 54896, Republic of Korea}
\affiliation[b]{Division of Science Education and Institute of Fusion Science, Jeonbuk National University, Jeonju 54896, Republic of Korea}
\affiliation[c]{Instituto de Física Corpuscular, CSIC-Universitat de València, Paterna 46980, Spain}
\affiliation[d]{Departament de Física Teòrica, Universitat de València, Burjassot 46100, Spain	
}

\abstract{
We show that, when connected with monopoles, the \textit{flat} $D$-flat direction breaking the local $U(1)_{B-L}$ symmetry as an extension of the minimal supersymmetric standard model can be responsible for the signal of a stochastic gravitational wave background recently reported by NANOGrav collaborations, while naturally satisfying constraints at high frequency band.
Thanks to the flatness of the direction, a phase of thermal inflation arises naturally.
The reheating temperature is quite low, and suppresses signals at frequencies higher than the characteristic frequency set by the reheating temperature.
Notably, forthcoming spaced based experiments such as LISA can probe the cutoff frequency, providing an indirect clue of the scale of soft SUSY-breaking mass parameter.
}

\emailAdd{rinkumaji9792@gmail.com}
\emailAdd{wipark@jbnu.ac.kr}
\makeatletter
\gdef\@fpheader{}
\makeatother
\begin{document}

\maketitle

\section{Introduction}
\label{sec:intro}

Recently, NANOGrav collaborations have reported a discovery of a stochastic gravitational wave background (SGWB) \cite{NANOGrav:2023gor}. Other pulsar timing array experiments have also reported evidence of the stochastic gravitational wave background \cite{Antoniadis:2023ott,Reardon:2023gzh,Xu:2023wog}.
Astrophysical origins such as supermassive blackhole binaries and cosmological sources have been discussed for the signal \cite{NANOGrav:2023hfp,NANOGrav:2023hvm,EPTA:2023xxk}. 
A number of new physics models have been proposed to explain the data \cite{Madge:2023cak,Franciolini:2023wjm,Shen:2023pan,Zu:2023olm,Lambiase:2023pxd,Han:2023olf,Guo:2023hyp,Wang:2023len,Ellis:2023tsl,Fujikura:2023lkn,Kitajima:2023cek,Li:2023yaj,Franciolini:2023pbf,Megias:2023kiy,Ellis:2023dgf,Bai:2023cqj,Yang:2023aak,Murai:2023gkv,Ghoshal:2023fhh,Deng:2023btv,Huang:2023chx,Lazarides:2023ksx,Athron:2023mer,Oikonomou:2023qfz,Blasi:2023sej,Datta:2023vbs,Basilakos:2023xof,Lazarides:2023rqf,Vagnozzi:2023lwo,Borah:2023sbc,Barman:2023fad,Broadhurst:2023tus,Liu:2023pau,Liu:2023ymk,Ghosh:2023aum,Niu:2023bsr,Zhu:2023lbf,Jiang:2023gfe,Li:2023bxy,Anchordoqui:2023tln,DiBari:2023upq,Li:2023tdx, Du:2023qvj,Antusch:2023zjk,Choudhury:2023kam,Ye:2023xyr,Yamada:2023thl,Zhang:2023nrs,Ben-Dayan:2023lwd,Cheung:2023ihl,Aghaie:2023lan,Ahmadvand:2023lpp,An:2023jxf,King:2023ayw,He:2023ado,Bhaumik:2023wmw,Ellis:2023oxs,Ahmed:2023rky}.
Among the variety of possibilities meta-stable cosmic strings associated with a symmetry-breaking scale close to that of Grand Unified Theory (GUT) might be a quite plausible source beyond the standard model since they can appear naturally in various GUT models \cite{Chakrabortty:2017mgi,Buchmuller:2019gfy,Buchmuller:2020lbh,Buchmuller:2021dtt,Buchmuller:2021mbb,Buchmuller:2023aus,Antusch:2023zjk,Lazarides:2023rqf}.
However, such a high scale symmetry-breaking predicts too large signal amplitude at high-frequency band to be consistent with constraints from observations \cite{LIGOScientific:2021nrg}.
The problem can be solved if the topological defects experience some amount of $e$-folding before the end of inflation, but the symmetry-breaking should happen at certain epoch roughly in the middle of about 30 $e$-foldings \cite{Maji:2022jzu,Lazarides:2023ksx,Lazarides:2023rqf}.
Alternatively, if the reheating temperature of inflation or an early matter-domination era due to certain long-lived particles is low enough, high frequency signal can be suppressed \cite{Cui:2018rwi}. 
The latter possibility can be easily realized in supersymmetric theories in which symmetry-breaking flat-directions are regarded as natural entities.
 
It is well-known that, if there is a flat symmetry-breaking field which has large enough coupling to the thermal bath at the symmetric phase, there can be an epoch of thermal inflation which can erase pre-existing dangerous relics by a huge amount of late-time entropy injection \cite{Lazarides:1986rt,Lyth:1995hj,Lyth:1995ka}.
It is actually the most compelling solution so far to the cosmological moduli problem \cite{Coughlan:1983ci,Ellis:1986zt,deCarlos:1993wie,Banks:1993en,Randall:1994fr,Bento:1994qg,Banks:1995dt} which is a quite robust generic problem in supergravity or string theories.
In supersymmetric theories, the flatness of a potential can have a direct link to the soft SUSY-breaking mass parameter, $m_s$ of which we are eager to get information in any available experiments.
If the potential causes thermal inflation, the mass scale combined with a large symmetry-breaking vacuum expectation value close to GUT scale can end up a quite low reheating temperature.
Hence, if such a flat field is responsible for the formation of strings, the gravitational waves (GWs) from such strings can have a quite low high-frequency cutoff of the signal, easily satisfying observational constraints at high frequency band.
If the impact of such a cutoff on the GW signal is probed at forthcoming experiments, such as LISA, it may be possible to get a clue of SUSY indirectly, although it might be non-trivial to pin down the model among various other possibilities.

Strings formed from a flat symmetry-breaking potential have a couple of distinctive features.
The core width of field-theoretic strings is given by roughly the inverse of the mass scale of the string-forming scalar field.
The width determines the characteristic length scale of strings below which the emission of particles dominates \cite{Blanco-Pillado:1998tyu,Olum:1998ag,Matsunami:2019fss,Auclair:2019jip}, which plays the role of a high frequency cutoff.
For critical or Type-II Abelian-Higgs strings considered typically, the mass scale is similar to or larger than that of the associated gauge field.
It is roughly of around the symmetry-breaking scale, and the associated frequency-cutoff of the GW signal is located at a high-frequency region well above current or near-future experimental reaches. 
On the other hand, the width of strings from a flat potential is much larger than that of critical or Type-II strings.
Especially, in supersymmetric theories the mass of the scalar field associated with a flat potential is typically of the scale of soft SUSY-breaking mass parameter, $m_s$ which might be of $\mathcal{O}(10-100) {\rm TeV}$.
It is smaller than the GUT-like symmetry-breaking scale by many orders of magnitude.
Hence the high-frequency cutoff can be lowered down by several orders of magnitude compared to that of critical or Type-II strings having the same symmetry-breaking scale.
Also, strings from a flat potential feel attractive force among themselves irrespective of their orientations as they get closer to start overlapping \cite{Bettencourt:1994kc}.
This attractive force is known to cause so-called zippering effects, which allow the formation of higher winding states \cite{Cui:2007js}. 
The effect results in enhancements of GW signals, and becomes stronger toward lower frequencies \cite{Jeong:2023iei}.
 
Motivated by these aspects of flat potential, in this paper we propose a scenario of meta-stable cosmic strings associated with a scalar field responsible for thermal inflation.
Especially, we consider a supersymmetric local $U(1)_{B-L}$ extension of the minimal supersymmetric standard model, discussing possibilities of connecting to monopoles that can arise from earlier symmetry-breaking processes.
The structure of the paper will be as follows.
In Section~\ref{sec:model}, the model is briefly described.
In Section~\ref{sec:thermal_infl}, thermal inflation and related cosmological aspects are briefly discussed.
in Section~\ref{sec:GWs}, an estimation of the expected gravitational wave signal in our scenario is shown in comparison to the NANOGrav data.
In Section~\ref{sec:concl}, a summary and conclusions are drawn.

\section{The Model}
\label{sec:model}

Topological defects in field theories depend on the nature of the vacuum manifold ($\mathcal{M}$).
Depending on theories, there are a variety of possibilities classified by homotopy group $\pi_n(\mathcal{M})$.
Monopoles or strings can form if $\pi_2(\mathcal{M}) \neq I$ or $\pi_1(\mathcal{M}) \neq I$, respectively.
Hybrid defects such as monopoles connected by string are formed if a symmetry group $G$ breaks down as $G \to H \to K$ with 
\beq
\pi_2(G/H) \neq I, \quad \pi_1(H/K) \neq I\quad \mathrm{but} \quad \pi_1(G/K) = I
\eeq
Depending on model parameters, they can live long enough, leaving some observable imprints such as GWs in the present universe. 

In this work we are interested in GUT models in which $U(1)_{B-L}$ is contained and broken along flat directions at a temperature low but well above the electroweak scale.
We pay attention to the case in which a cosmic string connects a monopole to an antimonopole, forming meta-stable strings which can be responsible for the GW signal recently discovered by PTA collaborations across the world. Examples of realizing the composite defects of monopoles and strings were discussed in Refs.~\cite{Lazarides:2019xai,Lazarides:2023iim,Buchmuller:2023aus}.
A simple structure of UV theories for such a possibility can have the following pattern of symmetry-breaking:
\bea
&&SU(3)_c \times SU(2)_L \times U(2) \ \l( U(2) = SU(2)_{\rm R} \times U(1)_{B-L}/\mathbb{Z}_2 \r)
\nonumber \\
&\xrightarrow[]{M_R}& SU(3)_c \times SU(2)_L \times U(1)_R \times U(1)_{B-L} 
\nonumber \\
&\xrightarrow[]{M_{BL}}& SU(3)_c \times SU(2)_L \times U(1)_Y
\eea
We take it as a UV-example for our low-energy effective theory.
Another UV model can be based on $SU(4)_c\times SU(2)_L\times U(1)_R$ gauge group, and the models can be embedded in Pati-Salam model \cite{Pati:1974yy} and the GUT model $SO(10)$ \cite{Fritzsch:1974nn}.
As discussed in Ref.~\cite{Buchmuller:2023aus}, we assume that there are one $SU(2)_R$ triplet neutral under $U(1)_{B-L}$ and two $B-L$ charged $SU(2)_R$ doublet Higgses.
The former is responsible for the breaking of $SU(2)_R$ and formation of monopoles, and and the latter breaks $U(1)_R\times U(1)_{B-L}$ eventually, leading to the formation of meta-stable strings.
As a concrete UV-realization, we consider a simple model obtained from the one in Eq.~(2.26) of Ref.~\cite{Buchmuller:2023aus} by setting $\lambda = 0$ and $h = 0$.
Setting $\lambda = 0$ allows the light $D$-flat direction consisting of $SU(2)_R$ doublets.
In this circumstance, the associated symmetry-breaking can be realized due to the negative soft SUSY-breaking mass-squared of the flat direction.
Setting $h=0$ is harmless, since the $U(1)_Y$ preserving symmetry-breaking can be the one selected anthropically. 
Also, we assume that monopoles formed from the triplet Higgs were inflated away by the primordial inflation.
 
The low-energy effective theory in our scenario can be described by the following supersymmetric extension of the MSSM invariant under $SU(3)_c \times SU(2)_L \times U(1)_R \times U(1)_{B-L}$,
\bea \label{eq:W-full}
W 
&=& W_{{\rm MSSM}-\mu} + \mu_H H_uH_d + \mu_\Phi \Phi_1 \Phi_2 + y_\nu LH_u N
\nonumber \\ 
&& + \frac{\lambda_N}{M} \Phi_1^2 N^2 + \frac{\lambda_H}{M} \l( H_u H_d \r)^2 + \frac{\lambda_\mu}{M} \Phi_1 \Phi_2 H_u H_d + \frac{\lambda_\Phi}{M} \l( \Phi_1 \Phi_2 \r)^2
\eea 
where $W_{{\rm MSSM}-\mu}$ includes the conventional MSSM Yukawa terms, $N$ is the right-handed neutrino superfield, $\Phi_1$ and $\Phi_2$ are originated from the $B-L$ charged $SU(2)_R$ doublet Higgs chiral superfields, carrying $U(1)_{B-L}$ charges as 
\beq
q_{BL} (\Phi_1, \Phi_2) = (1, -1),
\eeq
and $M$ is the cutoff scale of the effective interactions and taken to be of GUT scale for definiteness.
The dimensionless coupling constant $\lambda_i$s are taken as free parameters.
This model is essentially the same as the one considered in Ref.~\cite{Jeong:2023iei} except the fact that here the mass-term of right-handed neutrino is realized by a non-renormalizable higher order operator.
The physics involved and cosmological aspects are more or less the same, except that we now take a look at the case of GUT-scale symmetry-breaking, which is close to the upper-bound of the allowed range of the VEVs of $\Phi_1$ and $\Phi_2$.
So, we refer the readers to Ref.~\cite{Jeong:2023iei} for various details of the model, but provide the form of the potential along the $D$-flat direction consisting of $\Phi_1$ and $\Phi_2$ for later convenience, 
\beq \label{eq:V-Dflat}
V
= V_0 + \frac{1}{2} \l( m_1^2 + m_2^2 \r) |\phi|^2
- \frac{1}{2} \l[ B_\Phi \mu_\Phi \phi^2 + \frac{A_\Phi \lambda_\Phi}{4M} \phi^4 + {\rm c.c.} \r] + \l| \mu_\Phi 
+ \frac{\lambda_\Phi \phi^2}{2 M} \r|^2 |\phi|^2
\eeq
where $m_i^2$ is the soft SUSY-breaking mass-squared parameters of $\Phi_1$ and $\Phi_2$, $B_\Phi$ and $A_\Phi$ are soft SUSY-breaking parameter associated with the $B-L$ bilinear and higher order terms in the superpotential of \eq{eq:W-full}, and the $D$-flat direction is denoted as $\phi$ with $\Phi_1 \equiv \phi/\sqrt{2} \equiv \Phi_2$. 
We assume $m_1^2+m_2^2 < 0$ so that the symmetry is broken along the direction.
Ignoring contributions involving $\mu_\Phi$ under the assumption of $\mu_\Phi \ll m_s$ for simplicity, the vacuum position, $\phi_0$, is found to be
\bea \label{eq:phi0}
\phi^2_0 \approx \frac{A_\Phi M}{3 \lambda_\Phi} \l[ 1 + \sqrt{1 + \frac{12 m_{\phi,0}^2}{A_\Phi^2}} \r],
\eea
where $m_{\phi,0}^2 \equiv - \l(m_1^2 + m_2^2 \r)/2 > 0$, and $CP$-conservation was assumed. 
For vanishing cosmological constant with $m_{\phi,0} \sim A_\Phi \sim B_\Phi$, \eq{eq:phi0} gives
\beq
V_0 \approx m_{\phi_0}^2 \phi_0^2
\eeq

\section{Thermal inflation}
\label{sec:thermal_infl}
In the framework of supergravity which we assume implicitly as the underlying framework, the effective mass-squared during inflation is expected to be
\cite{Gibbons:1977mu,Shafi:1983bd}
\beq
m_{\phi, \rm eff}^2 = -m_{\phi,0}^2 + c_H H_I^2
\eeq
where $c_H$ is a dimensionless parameter of order unity with either sign allowed and $H_I$ is the expansion rate during inflation.  
For $H_I \gg  m_{\phi,0} \sim m_s$, if $c_H>0$, it is possible to have $m_{\phi, \rm eff}^2 >0$, holding $\phi$ near the origin.
With this condition satisfied, thermal inflation can take place if the background temperature of the universe $T$ after inflation is large enough, i.e., $T \gg m_{\phi,0}$ so that $\phi$ can be held around the origin while $V_0$ dominates the energy density of the universe.

We assume for simplicity an efficient reheating after inflation so that the universe was dominated by radiation at the epoch of the onset of thermal inflation.
The presence of Planckian moduli causes only a minor change in this picture, so we ignore such details.
Thermal inflation begins at $T=T_{\rm b}$ which is given by
\beq
T_b \sim V_0^{1/4}
\eeq
It ends at $T=T_{\rm c} \sim m_{\phi,0}$ under the assumption of order unity coupling of $\phi$ to thermal bath.
The duration (or $e$-foldings) of thermal inflation is then given by
\beq
N_e^{\rm TI} \approx \ln \l( \frac{T_{\rm b}}{T_{\rm c}} \r) \sim \ln \l( \frac{V_0^{1/4}}{m_{\phi,0}} \r) \sim \frac{1}{2} \ln \l( \frac{\phi_0}{m_s} \r) \sim 13
\eeq 
where in the last approximation we used $m_{\phi,0} \sim m_\phi \sim m_s$ with $m_\phi$ being the physical mass of the flaton field $\phi$ at the true vacuum position $\phi_0$.

When thermal inflation ends along a flat direction with a large symmetry-breaking scale, there will be a long matter(flaton)-domination era until flaton decays.
In our scenario, the presence of $\lambda_\mu$ coupling allows efficient decays of flaton, leading to a decay rate given by \cite{Kim:2008yu},
\beq \label{eq:Gamma-phi}
\Gamma_\phi 
\approx \frac{1}{4 \pi} \frac{m_\phi^3}{\phi_0^2} \l( \frac{m_A^2 - |B|^2}{m_A^2} \r)^2 \l( \frac{|\mu|}{m_\phi} \r)^4
\sim \frac{1}{4 \pi} \frac{m_\phi^3}{\phi_0^2} \l( \frac{|\mu|}{m_\phi} \r)^4,
\eeq
where $\mu=\lambda_\mu \phi_0^2/2M$, the mass of the heavy CP-odd Higgs boson is given by 
\beq
m_A^2 \equiv m_{H_u}^2 + m_{H_d}^2 + 2 |\mu_{\rm eff}|^2,
\eeq
and $B$ denotes the soft-SUSY breaking parameter associated with the $\lambda_\mu$-term.
In \eq{eq:Gamma-phi}, we assumed that $m_\phi \gg m_h = 125 {\rm GeV}$, but smaller than twice of the mass of the neutralino LSP \footnote{If $R$-parity is violated so that the LSP is unstable, $m_\phi$ is not constrained by the mass of the LSP.}. 
The reheating temperature of thermal inflation (i.e., the temperature at the epoch of flaton's decay) is then found as \cite{Jeong:2023iei}
\beq
T_{\rm d} = \l( \frac{\pi^2}{45} g_*(T_{\rm d}) \r)^{-1/4} \sqrt{\frac{2}{5} \Gamma_\phi m_{\rm Pl}}.
\eeq
where $g_*(T_{\rm d})$ is the number of relativistic degrees of freedom at $T_{\rm d}$ and $m_{\rm Pl}$ is the reduced Planck mass.
As one can see, even for $\phi_0 \sim M_{\rm GUT}$ with $m_\phi \gtrsim \mathcal{O}(1) {\rm TeV}$, decay of the flaton $\phi$ is efficient enough so as to reheat the universe with the reheating temperature $T_{\rm d} \gtrsim \mathcal{O}(1-10) {\rm MeV}$ for a successful big-bang nucleosynthesis.

Thermal inflation is the best known solution to the cosmological moduli problem expected in supergravity and string theories that might be the underlying theories when theories beyond the standard model are extended to include supersymmetry.
However, it dilutes not only dangerous long-lived moduli or gravitinos but also pre-existing baryon number asymmetry.
Because of this reason, most baryogenesis mechanisms are incompatible with thermal inflation, and the genesis should be redone after the end of thermal inflation.
In our scenario, even though $\phi_0 \sim M_{\rm GUT}$ which can lead to a quite low reheating temperature, the late-time Affleck-Dine mechanism utilizing $\lambda_\mu$-term \cite{Stewart:1996ai,Jeong:2004hy,Kim:2008yu,Park:2010qd} still works as long as one of the $LH_u$ flat-direction is flat enough to develop large temporal VEV.
On the other hand, the best candidate for dark matter is likely to be QCD-axions or axion-like particles since all different types of pre-existing dark matter candidates are expected to be diluted out.

In our scenario, the end of thermal inflation is the $U(1)_R\times U(1)_{B-L}$-breaking phase transition. 
During this process, Type-I thick cosmic strings are formed.
As time goes on, depending on the mass $m_{\rm M}$ of the monopoles associated with the breaking of $SU(2)_R$ and the tension $\mu_s$ of the strings, monopole-antimonopole pairs can populate on strings, segmenting strings.
As a result, the string network can decay.
This decay provides a low-frequency cutoff of the GW signals from string loops and allow it to match the discovered signal as described in the next section.


\section{Stochastic gravitational wave backgrounds for NANOGrav 15 year data}
\label{sec:GWs}

The gravitational waves from composite defects have been discussed in many papers, including Refs.~\cite{Martin:1996ea,Martin:1996cp,Leblond:2009fq,
Buchmuller:2019gfy,Buchmuller:2020lbh,Blasi:2020wpy,
Buchmuller:2021dtt,Buchmuller:2021mbb,Masoud:2021prr,Dunsky:2021tih,Ahmed:2022rwy,Afzal:2022vjx,Lazarides:2022jgr,Yamada:2022imq,Yamada:2022aax,Maji:2023fba}.
The spectral pattern of the expected GW signal from cosmic strings depends on the stability of the strings.  
When the cosmic strings are not topologically stable, they can cut into pieces via the production of monople-antimonopoles on strings, and their decay rate per unit length in the semiclassical approximation is given by \cite{Preskill:1992ck,Monin:2008mp,Leblond:2009fq} \footnote{The core width of strings in our scenario is many orders of magnitude larger than that of monopoles.
It is unclear how significantly this fact would affect the decay rate of strings, and to our knowledge, there has been no work on this subject.
If the large core width causes a suppression of the decay rate, we would have to take a smaller monopole mass. 
As long as such an adjustment can work, our argument remains unaffected.}
\begin{align}\label{eq:gam_d}
\Gamma_s \approx\frac{\mu_s}{2\pi}\exp\left(-\pi\kappa\right),
\end{align}
where $\kappa=m_M^2/\mu_s$ with $m_M$ being the monopole mass, and the string tension which depends on the winding number $N_w$ is expressed as \cite{Cui:2007js}
\beq \label{eq:mu-Nw}
\frac{\mu_s(N_w)}{\pi \phi_0^2} 
\approx c_1 \l( 1 + c_2 \ln N_w \r)
\eeq
with the coefficients:
\bea \label{eq:coeff-c1}
c_1 &\approx& \frac{4.2}{\ln \l(R \r)} + \frac{14}{\ln^2 \l(R\r)}, 
\\ \label{eq:coeff-c2}
c_2 &\approx& \frac{2.6}{\ln \l(R \r)} + \frac{57}{\ln^2 \l(R\r)} .
\eea
where, $R={m_V^2}/{m_\phi^2}$ is the ratio of the squared mass of the vector to the scalar field associated with the symmetry breaking. 
For strings formed from a very flat potential, strings with different winding numbers are expected to appear due to zippering between the strings \cite{Cui:2007js}.
Those different string species seem to be equally distributed in terms of their number density.
Also, the maximum winding number achieved at an epoch grows as the universe evolves.
This effect causes a spectral change of the GW signal relative to the case without zippering effects.
A full consideration of the effect for a precise estimation requires a thorough numerical simulation. 
Instead, we qualitatively estimate its impact on the gravitational wave using a power averaged over winding numbers. 
The average power is estimated as \cite{Cui:2007js,Jeong:2023iei}
\beq \label{eq:P-average}
\overline{P}_{\rm GW}(t)
\xrightarrow[]{N_w^{\rm max} \gg 1} \Gamma G \mu_s^2(N_w^{\rm max}(t)) .
\eeq
 Here, $\Gamma \simeq 50$ is a numerical factor obtained from simulations \cite{Vachaspati:1984gt, Vilenkin:2000jqa} and
$N_w^{\rm max}$ at a time $t$ is numerically found in Ref.~\cite{Cui:2007js}:
\beq \label{eq:NwMax}
N_w^{\rm max}(t) \sim \l( \frac{t}{t_c} \r)^{0.22},
\eeq
where  $t_c$ is the time when thermal inflation ends. 

We assume that the gravitational wave bursts from cusps dominate and the waveform is given by \cite{Damour:2001bk}
\begin{align}
h(f,l,z) = g_{1c}\frac{G\mu_{s,c} l^{2/3}}{(1+z)^{1/3}{r(z)}}f^{-4/3},
\end{align}
where $g_{1c}\simeq 0.85$ \cite{LIGOScientific:2021nrg}, $\mu_{s,c} \equiv \mu_s(N_w=1)$, $l$ denotes the loop length, $z$ is the cosmological redshift, and $r(z)$ is the proper distance. The rate of bursts is given by
\begin{align}\label{eq:burst-rate}
\frac{d^2R}{dz \, dl} = N_c \l( 1 + c_2 \ln N_w^{\rm max}(t) \r)^2 \frac{4\pi r^2(z)}{(1+z)^3H(z)} \frac{2n(l,t(z))}{l(1+z)}\left( \frac{\theta_m(f,l,z)}{2}\right)^2\Theta(1-\theta_m),
\end{align}
 where $N_c=2.13$ \cite{Cui:2019kkd} is the average number of cusp per winding on a loop, $H$ is the Hubble parameter, and $\theta_m$ is the beam opening angle given by
\begin{align}
\theta_m(f,l,z) = \left[\frac{\sqrt{3}}{4}(1+z)fl\right]^{-1/3}.
\end{align} 
The loop distribution $n(l,t)$ in a radiation dominated universe is expressed as \cite{Blanco-Pillado:2013qja,Blanco-Pillado:2017oxo,LIGOScientific:2021nrg,Buchmuller:2021mbb,Buchmuller:2023aus}
\begin{align}\label{eq:n-loop-rad}
n(l,t<t_s) &=
 \frac{0.18 \ \Theta(0.18t-l)}{t^{3/2}(l+\Gamma G\mu_{s,c} t)^{5/2}} , \\
n(l,t>t_s) &= \frac{0.18\ \Theta(0.18t-l-\Gamma G\mu_{s,c}(t-t_s))}{t^{3/2}(l+\Gamma G\mu_{s,c} t)^{5/2}}\exp\left[-\Gamma_d\left( l(t-t_s)+\frac{1}{2}G\mu_{s,c}(t-t_s)^2\right)\right] 
\end{align}
and that during early matter era will be given as
\begin{align}\label{eq:n-loop-mat}
n_{m}(l,t) = \frac{0.27-0.45(l/t)^{0.31}}{t^2(l+\Gamma G\mu t)^2}\Theta(0.18t-l) ,
\end{align}
where $t_s=1/\sqrt{\Gamma_s}$. 
Early matter domination era continues from $t_c$ till $t_d=\frac{5}{4}\Gamma_\phi^{-1}$. 

 Equipped with this, we estimate the gravitational wave background from the flat-direction cosmic strings using the burst method given by \cite{Damour:2001bk,Olmez:2010bi, Auclair:2019wcv, Cui:2019kkd, LIGOScientific:2021nrg}
\begin{align}\label{eq:GWs-Omega-cusps}
\Omega_{GW}(f) = \frac{4\pi^2}{3H_0^2}f^3\int_{z_m}^{z_F} dz \int dl \, h^2(f,l,z)\frac{d^2R}{dz \, dl} ,
\end{align}
where  $z_m$ is set to separate the contributions of the infrequent bursts from the stochastic background:
\begin{align}
f = \int_{0}^{z_m} dz \int dl \frac{d^2R}{dz \, dl}
\end{align}
and $l$ can be taken from $0$ to $2t$ ($3t$) in the radiation (matter) dominated era. However, various theta functions in Eqs.~\eqref{eq:burst-rate}, \eqref{eq:n-loop-rad} and \eqref{eq:n-loop-mat} select appropriate limits of the integration. The upper limit of integration for the redshift $z$ in our case will be $z_F=\mathrm{min}[z(t_c), z(t_p)]$ where $t_p=1/(\alpha m_\phi\Gamma G\mu)$ is the timescale upto which the  particle emission dominates \cite{Cui:2007js}. 
\begin{figure}[htbp!]
\begin{center}
    \includegraphics[width=0.75\linewidth]{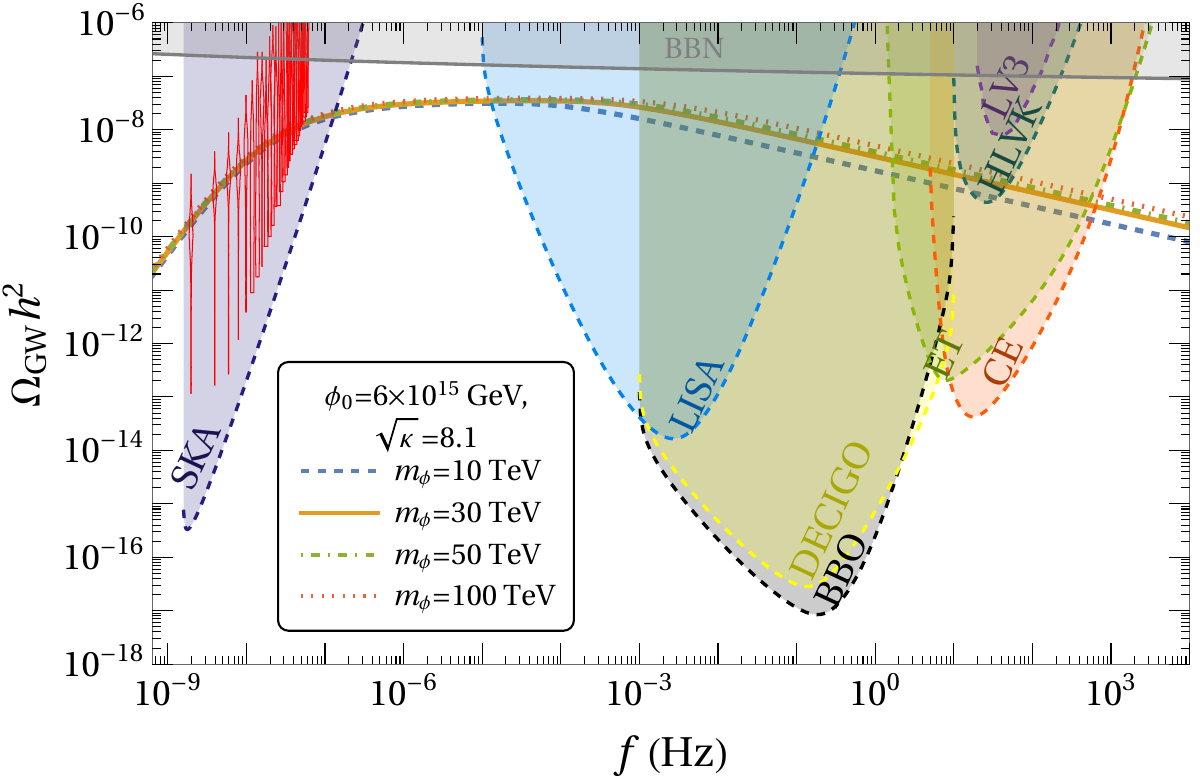}
\end{center}
\caption{Gravitational wave background spectra from flat-direction metastable cosmic strings with $\phi_0 = 6\times 10^{15}$ GeV for varying values of the mass parameter $m_\phi \in [10,100]$ TeV. The red violin plots dictate the HD correlated free spectra of NANOGrav 15 year data.
Big Bang Nucleosynthesis (BBN) constraint \cite{Mangano:2011ar} is also depicted.
The gravitational wave spectra can explain the NANOGrav 15 year data for $t_s\approx 3\times 10^5$ sec and are consistent with the advanced LIGO-VIRGO third observing run (LV3) bound \cite{LIGOScientific:2021nrg}. 
%
They are within reach of several proposed experiments shown as the power-law integrated sensitivity curve \cite{Thrane:2013oya, Schmitz:2020syl} of each experiment named in the figure. 
}
\label{fig:GWs-MSS}
\end{figure}

Fig.~\ref{fig:GWs-MSS} shows the stochastic gravitational wave background from flat-direction metastable cosmic strings with $\phi_0 = 6\times 10^{15}$ GeV for varying values of the mass parameter $m_\phi$ within $[10,100]$ TeV. 
The expected signal at low frequency region is insensitive to $m_\phi$.
This is because $m_\phi$-dependence is mainly from the string tension, but only weakly through the log-factors in \eqs{eq:coeff-c1}{eq:coeff-c2}.
The critical impact of $m_\phi$ is on the bending frequency where the spectrum turns from a plateau to $f^{-1/3}$ behavior.
On the other hand, $\phi_0$-dependence is much stronger, and roughly given by $\Omega_{\rm GW} \propto \phi_0$ if the logarithmic dependence in the string tension is ignored.  
From the figure, one can see that the gravitational wave spectra expected in our scenario can explain the NANOGrav 15 year data for $\sqrt{\kappa}=8.1$ and are consistent with the advanced LIGO-VIRGO third observing run (LV3) bound \cite{LIGOScientific:2021nrg}. The expected spectra of gravitational waves can be tested in many proposed experiments, including HLVK \cite{KAGRA:2013rdx}, LISA \cite{Bartolo:2016ami, amaroseoane2017laser}, CE \cite{Regimbau:2016ike}, ET \cite{Mentasti:2020yyd}, DECIGO \cite{Sato_2017}, BBO \cite{Crowder:2005nr, Corbin:2005ny} and SKA \cite{5136190, Janssen:2014dka}. 

It is worth mentioning that the gravitational spectra consist of a $f^2$ rising part in the low frequencies \cite{Buchmuller:2021mbb}, a (nearly) scale-invariant plateau in the intermediate frequency, and a falling $f^{-1/3}$ behavior in the high-frequency region.
The spectral bending to $f^{-1/3}$ behavior appears at frequency $f_*$ determined by the reheating temperature, $T_{\rm d} \sim \sqrt{\Gamma_\phi m_{\rm Pl}} \propto m_\phi^{3/2}/\phi_0$.
For the parameter sets taken in the figure, we have  $T_{\rm d} \sim (0.1-1.0)~\mathrm{GeV}$ for $|\mu|=m_\phi$ in the approximation in \eq{eq:Gamma-phi}. 
It leads to $f_* \sim (10^{-4}-10^{-3})~\mathrm{Hz}$ located within the LISA sensitivity band.

We perform a Bayesian analysis with the NANOGrav 15 year data set using the package \texttt{PTArcade} \cite{Mitridate:2023oar}. We use \texttt{Ceffyl} \cite{Lamb:2023jls} to sample the posterior distributions of the model parameters taking the Hellings-Downs (HD) correction \cite{Hellings:1983fr} among the pulsars for the gravitational wave into account.
\begin{figure}[htbp!]
\begin{center}
\includegraphics[scale=0.9]{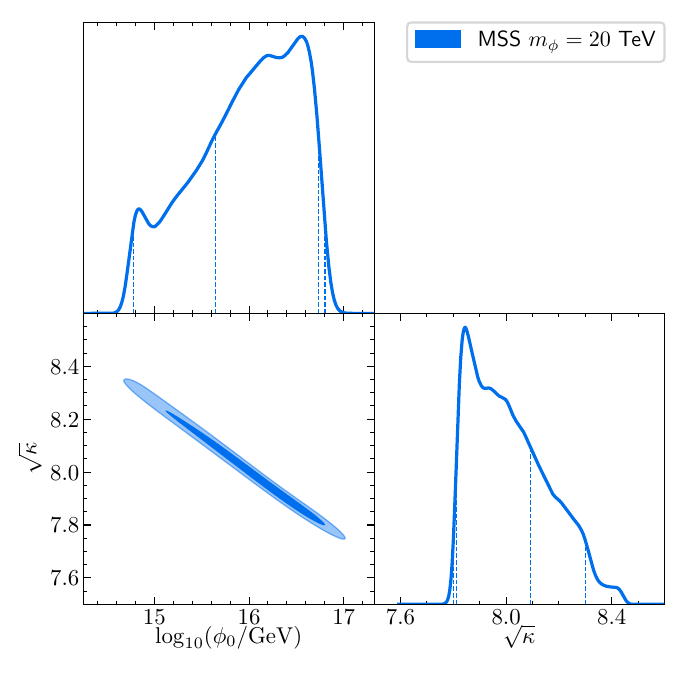}
\end{center}
\caption{Triangular plot of the posterior distribution of the model parameters $\log_{10}(\phi_0/\mathrm{GeV})$ and $\sqrt{\kappa}$ for $m_\phi=20$ TeV with $1\sigma$ (dark blue) and $2\sigma$ (light blue) Bayesian credible regions.}\label{fig:bayes}
\end{figure}
\begin{table}
\centering
\begin{tabular}{lcccc}
\hline
 \multirow{2}{*}{Parameters} & \multicolumn{2}{c}{Bayesian Credible Intervals} \\
\cline{2-3}
  & 68\% & 95\% \\
\hline
 $\log_{10}(\phi_0/\mathrm{GeV})$ & $[15.64, 16.73]$ & $[14.78, 16.80]$ \\
 $\sqrt{\kappa}$ & $[7.81, 8.09]$ & $[7.80, 8.30]$ \\
\hline
\end{tabular}
\caption{Bayesian credible intervals for the parameters $\phi_0$ and $\kappa$ in the model with $m_\phi=20$ TeV.}
\label{tab:params}
\end{table}
Fig.~\ref{fig:bayes} shows the corner plot of the model parameters $\phi_0$ and $\kappa$ and Table~\ref{tab:params} dictates the $68\%$ and $95\%$ confidence level intervals. We have taken $m_\phi=20$ TeV without loss of the generality around the nanoHertz frequencies (see Fig.~\ref{fig:GWs-MSS}).

\section{Conclusions}
\label{sec:concl}
In this work, we show that flat $B-L$ breaking supersymmetric $D$-flat direction can be responsible simultaneously for thermal inflation and the signal of stochastic gravitational waves observed by PTA collaboration across the world when it is embedded into a larger gauge group in which magnetic monopoles are expected.  The model with the parameters $\log_{10}(\phi_0/\mathrm{GeV})\in [14.8, 16.8]$ and $\sqrt{\kappa}\in [7.8, 8.3]$ can explain the evidence of stochastic gravitational wave in NANOGrav 15 year data.

The basic idea for the signal is the formation of meta-stable strings.
However, this work is different from earlier trials to address the observations because the scalar potential responsible for the formation of cosmic strings is quite flat.
Such a flatness has two-folded benefit in regard of the expected SGWBs in addition to realizing thermal inflation: 
\begin{itemize}
\item[(i)] High-frequency cutoff of the expected SGWBs at sub-${\rm Hz}$ region due to the long matter-domination era after thermal inflation with a quite low reheating temperature -
Existing constraint on the string tension at the frequency band of 100 Hz from advanced LIGO-VIRGO third observing run \cite{LIGOScientific:2021nrg} can be satisfied in a natural manner.
\item[(ii)] Enhancement of signals due to the zippering effect, which is expected to appear for strong Type-I strings (i.e. the mass of the scalar field is smaller than that of the gauge field by many orders of magnitude) -
This effect allows for a string network from the flat-potential to match the observed signal with GUT-scale symmetry breaking; otherwise, a larger breaking scale is required. 
\end{itemize}

The expected GW signal has a high-frequency spectral bending to $f^{-1/3}$ behavior caused by the low reheating temperature after thermal inflation.
Once probed by either LISA or DECIGO, it would provide an indirect hint of the scale of the soft SUSY-breaking mass-squared parameter.
Such bending can be caused by other physics, such as intermediate inflation without subsequent matter-domination era or effects of the small-scale structure of strings.
However, depending on the position of bending, there can be a strong preference for one of the various possibilities.


\medskip


\medskip


\section*{Acknowledgments}

\noindent
R.M. acknowledges Qaisar Shafi and Joydeep Chakrabortty for useful discussions.
W.I.P. acknowledges Gabriela Barenboim for helpful discussions at an early stage of this work.
This work was supported by the National Research Foundation of Korea grants by the Korea government: 2017R1D1A1B06035959 (W.I.P.) and 2022R1A4A5030362 (R.M and W.I.P.).
It was also supported by the Spanish grants PID2020-113775GB-I00 (AEI/10.13039/501100011033) and CIPROM/2021/054 (Generalitat Valenciana) (W.I.P.).

\bibliographystyle{JHEP}
\bibliography{GUT_TD_wip}

\end{document}